\documentclass[sn-mathphys]{sn-jnl}% Math and Physical Sciences Reference Style
\usepackage{graphicx}
\usepackage[numbers]{natbib}
\usepackage{color,soul}
\usepackage{multirow}
\usepackage{caption}
\usepackage{graphicx}	% A package for graphics use (see figures)
\usepackage{amsmath,amssymb}
\usepackage{adjustbox}
\jyear{2022}%

\theoremstyle{thmstyleone}%
\theoremstyle{thmstyletwo}%

\theoremstyle{thmstylethree}%
\newcommand{\com}[1] {}
\newcommand{\hlipcai}[1] {{1}}

\begin{document}

\title[Self-KD for Surgical Phase Recognition]{Self-Knowledge Distillation for Surgical Phase Recognition}


\author[1]{\fnm{Jinglu} \sur{Zhang}}\email{jinglu.zhang@medtronic.com}
\equalcont{These authors contributed equally to this work.}

\author[1]{\fnm{Santiago} \sur{Barbarisi}}\email{santiago.barbarisi@medtronic.com}
\equalcont{These authors contributed equally to this work.}

\author*[1]{\fnm{Abdolrahim} \sur{Kadkhodamohammadi}}\email{rahim.mohammadi@medtronic.com}

\author[1,2]{\fnm{Danail} \sur{Stoyanov}}\email{danail.stoyanov@medtronic.com}

\author[1]{\fnm{Imanol} \sur{Luengo}}\email{imanol.luengo@medtronic.com}

\affil[1]{\orgname{Medtronic Digital Surgery}, \orgaddress{\street{230 City
Road}, \city{London}, \country{UK}}}

\affil[2]{\orgdiv{Wellcome/EPSRC Centre for Interventional and Surgical
Sciences}, \orgname{University College London}, \city{London}, \country{UK}}

%%================================%%
%% Sample for structured abstract %%
%%================================%%

\abstract{\textbf{Purpose:} Advances in surgical phase recognition are generally led by training deeper networks. Rather than going further with a more complex solution, we believe that current models can be exploited better. We propose a self-knowledge distillation framework that can be integrated into current state-of-the-art (SOTA) models without requiring any extra complexity to the models or annotations.\\
\textbf{Methods:} Knowledge distillation is a framework for network regularization where knowledge is distilled from a teacher network to a student network. In self-knowledge distillation, the student model becomes the teacher such that the network learns from itself. Most phase recognition models follow an encoder-decoder framework. Our framework utilizes self-knowledge distillation in both stages. The teacher model guides the training process of the student model to extract enhanced feature representations from the encoder and build a more robust temporal decoder to tackle the over-segmentation problem.\\
\textbf{Results:} We validate our proposed framework on the public dataset Cholec80. Our framework is embedded on top of four popular SOTA approaches and consistently improves their performance. Specifically, our best GRU model boosts performance by \textbf{$+3.33\%$} accuracy and \textbf{$+3.95\%$} F1-score over the same baseline model.\\
\textbf{Conclusion:}
We embed a self-knowledge distillation framework for the first time in the surgical phase recognition training pipeline. Experimental results demonstrate that our simple yet powerful framework can improve performance of existing phase recognition models. Moreover, our extensive experiments show that even with 75\% of the training set we still achieve performance on par with the same baseline model trained on the full set.
}




\maketitle
\section{Introduction}\label{sec:intro}
With the increasing effort in improving patient outcomes and enhancing human-computer interaction, more attention is now being paid to develop context-aware surgical (CAS) systems for the operation room~\cite{maier2017surgical}. Surgical phase recognition is one of the fundamental tasks in developing CAS systems, which aims at dividing a surgical procedure into segments intra-operatively with reference to its pre-defined standard steps. Applications of real time phase recognition include surgery monitoring, resource scheduling, and decision making support. However, this task is challenging due to high intra-phase and low inter-phase variance, and the long duration of surgical videos. 

Some early studies used different variants of Hidden Markov Models (HMM)~\cite{padoy2008line} or statistic models~\cite{charriere2017real} to model phase transitions. Although these models are interpretable, for one, the handcrafted features have limited ability to represent high-dimensional surgical data. For another, these methods can only process a few local contiguous frames, thus struggling with capturing long-term temporal information. In recent decades,  deep neural networks (DNNs) have become the mainstream for surgical phase recognition by showing a superior ability to learn high-level features. 

EndoNet~\cite{cholec80} was the first phase recognition model that utilized a DNN (AlexNet~\cite{imagenet}) as its feature encoder and combined with support vector machine and HMM to detect surgical phases. With the rise in popularity of Recurrent Neural Networks (RNNs), DNNs are used both as encoder to extract latent features and temporal decoders to replace conventional graphical models. Most recent papers advanced phase recognition models by using deeper encoders, such as ResNet in TeCNO~\cite{tecno} and video Swin-transformer in GRU~\cite{intuitive_paper}, and sophisticated temporal decoders, for example temporal convolutional-based TeCNO~\citep{tecno} and Transformer-based Trans-SVNet~\cite{transsvnet}. Designing more advanced models is important for more reliable phase recognition. However, the questions that arise here are: \textit{Do we utilize the full capacity of these models? And should we always go for more complex solutions?}

We believe we can take advantage of knowledge distillation technique as a model-agnostic solution to improve the performance of surgical phase recognition. Knowledge distillation (KD)~\cite{hinton2015distilling} is a process of transferring knowledge from a heavier and better performing teacher model to a smaller one. Recent studies~\cite{kim2021self} regularize the model by distilling the internal knowledge from the student network itself, namely self-knowledge distillation (self-KD). The self-KD framework can be combined with various backbone models without increasing network complexity.

\begin{figure}[tb!]
    \centering
    \includegraphics[width=1\textwidth]{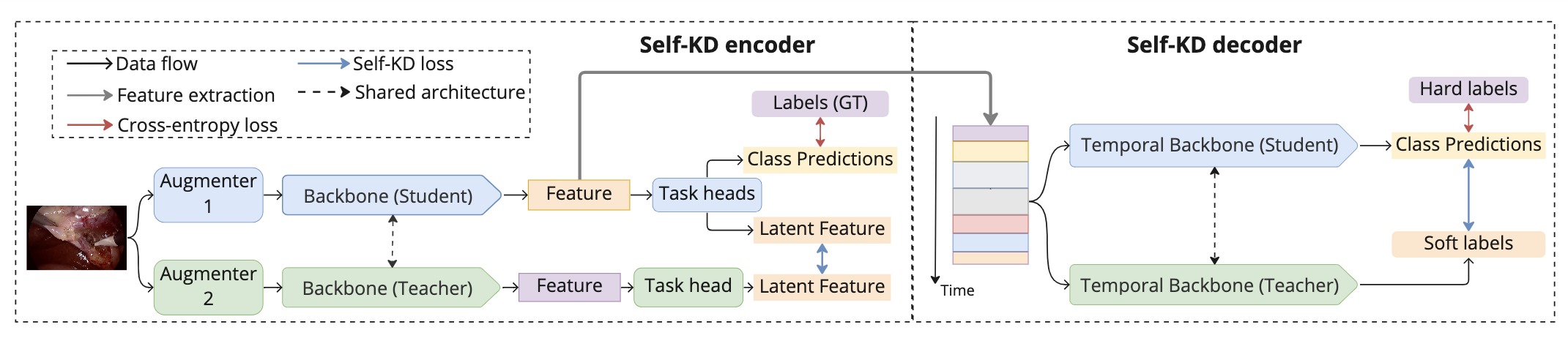}
    \caption{An overview of our self-knowledge distillation (self-KD) framework for surgical phase recognition. Our framework is composed of a self-KD encoder and a self-KD decoder. Self-KD encoder jointly optimizes a classification loss and feature similarity loss to enhance feature representation. While the self-KD decoder use the soft labels generated by the teacher model to regularize the prediction.}
    \label{fig_overview}
\end{figure}

\begin{figure}[t!]
    \centering
    \includegraphics[width=0.9\textwidth]{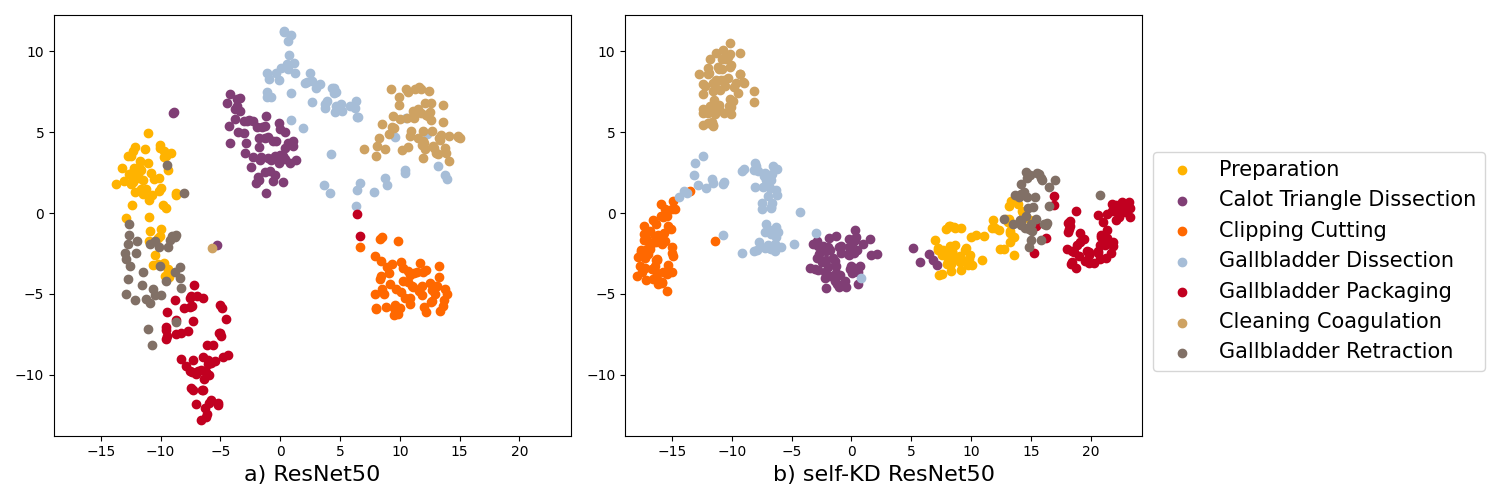}
    \caption{A 2D projection of video frame features computed using t-SNE: (a) baseline ResNet50; (b) self-knowledge distillation ResNet50. One can observe that \textit{Calot Triangle Dissection} or \textit{Gallbladder Packaging} obtained denser and more separable representations with self-knowledge distillation, hence facilitating the detection of those phases. }
    \label{fig:tsne}
\end{figure}

Inspired by these observations, we propose a novel self-knowledge distillation framework built upon existing phase recognition models without adding extra trainable modules or additional annotations. Our proposed framework integrates self-KD into both encoder and decoder. At the encoder stage, we train encoders by optimizing both frame classification and feature similarity objectives. This is in contrast with recent self-supervised approaches like SimCLR~\cite{simclr} or BYOL~\cite{byol}, that add an extra pre-training step where self-supervision is the only objective. In these methods, self-supervised approaches are applied to large general domain datasets~\cite{imagenet} where inter-phase variability is high. While in laparoscopic videos, this is not the case and inter-phase similarity is high. In our framework, frame representations are generated by the backbone of the student and forwarded into two task-specific heads for objective optimization (see self-KD encoder in~\autoref{fig_overview}). One head optimizes the classification objective similar to Czempiel et al. and Kadkhodamohammadi et al.~\cite{patg, tecno}. Concurrently, the second head optimizes a feature similarity objective. Inspired by~\cite{byol}, the target signal is generated by an auxiliary teacher\footnote{Student and teacher networks share the same architecture.} network. Moreover, student and teacher representations are obtained from differently augmented views of the same image. This enables us to obtain enhanced representations that are more robust to intra-phase variations. \autoref{fig:tsne} shows a low-dimension visualization of the features for different classes in a video. One can notice that self-KD helped the model to build more compact and separable representations.

At the decoder stage, we use self-KD to reduce the common over-segmentation problem~\cite{mstcn,patg} in temporal models. For a temporal backbone model, we take its best model from preceding epochs as the teacher model to generate soft labels. The soft labels from the teacher model force the student model (current epoch) to produce more consistent predictions for the same frame with variant logits. The model is regularized by minimizing teacher and student logits using a smoothing loss.

Following SOTA approaches for both phase recognition \cite{tecno, transsvnet} and self-supervision \cite{byol}, we benchmark our framework by training a ResNet50~\cite{resnet} encoder. As for the backbone decoder, we select four popular and representative temporal models as our backbone, including RNN-based (GRU~\cite{intuitive_paper}), temporal convolutional-based (TCN~\cite{tcn} and TeCNO~\cite{tecno}), and transformer-based (Trans-SVNet~\cite{transsvnet}) models. We validate the effectiveness of our approach on Cholec80~\cite{cholec80} dataset. The self-KD framework shows consistent improvement over four phase recognition models. Our contribution is two-fold:
\begin{itemize}
    \item To the best of our knowledge, we are the first to embed self-KD into surgical phase recognition architectures. It is a plug and play module. As a result, it can serve as a basic building block for any phase recognition model due to its generality and simplicity.
    \item With the teacher-student configuration, our method enables the encoder to extract enhanced spatial features and smooth predictions at the decoder. The extensive experiments (see~\autoref{sec:ab2}) demonstrate the robustness of our model with less training data. 
\end{itemize}

\section{Method}
The architecture of our self-knowledge distillation framework (see \autoref{fig_overview}) for surgical phase recognition is detailed in this section, and consists of two substructures: 1) A self-KD encoder (see \autoref{sec:encoder}) to extract representative features; 2) A self-KD decoder (see \autoref{sec:decoder}) integrated into existing models for reliable phase recognition. In general, a teacher-student structure is used, where a teacher network guides the student model. Both networks share the same structure, allowing the teacher to update the weights from the student, rather than optimizing them through back propagation. 
 
\subsection{Self-knowledge Distillation Encoder}\label{sec:encoder}
Encoders in phase recognition architectures obtain frame-wise feature representations from their last average pooling layer. Features from a given video are later forwarded into a temporal decoder for final predictions. The self-KD encoder aims to generate enhanced representations. In the self-KD encoder, a student model generates two outputs from an augmented view $x_{i,1}$ of an image $i$: frame-based phase probabilities $\hat{y}_{i,1}$ and a latent feature representation $z_{i,1}$ (see \autoref{fig_overview}). An auxiliary teacher model generates a second feature representation $z_{i,2}$ from the same image $i$ under a different augmentation $x_{i,2}$.

We optimize the student network by training two objectives simultaneously: phase recognition and feature similarity. Similar to \cite{tecno, patg} we use categorical cross-entropy loss $\mathcal{L}_{CE}$ to optimize over the phase probabilities $\hat{y}_{i,1}$. The similarity objective is optimized using mean-square error loss $\mathcal{L}_{MSE}$ over the normalized\footnote{Authors in \cite{simclr} show empirically that using a normalized projection improves model learning towards the downstream task.} feature representations from the student $\bar{z}_{i,1}$ and the teacher $\bar{z}_{i,2}$. Following a batch of $n$ images, the two losses are combined with equal weights into a single loss $\mathcal{L}^{E}_{self-KD}$ defined as:

\begin{equation}
    \mathcal{L}^{E}_{self-KD} = \mathcal{L}_{CE} + \mathcal{L}_{MSE} = - \sum_{i=0}^n \log (\hat{y}_i) + \sum_{i=0}^n \|\bar{z}_{i,1}-\bar{z}_{i,2} \|^2_2
\end{equation}

We do not back-propagate the loss through the teacher model. Rather, it is updated through Exponential Moving Average (EMA)~\cite{ema}, using parameters of the student model. Which means that the weights of the teacher model are a slow-moving average from the student. The teacher model is only used during training and discarded afterward.

\subsection{Self-knowledge Distillation Decoder}\label{sec:decoder}
The phase recognition decoder incorporates temporal dependencies to generate temporally smooth prediction over a video. The self-KD phase decoder aims to achieve more reliable predictions by regularizing the predictions from the backbone model. For any temporal model, we use its past predictions as an additional self-supervision signal (i.e. teacher model). The model is progressively regularized by minimizing the logits distance between student (the current predictions) and teacher model. 

For an input feature sequence $ x\in \mathbf{X}$ with length $L$: $x_{1:L} = (x_1,...,x_L)$, we aim to assign a class label $c \in \mathbf{C}$ to each frame: $c_{1:L}=(c_1,...,c_L)$. A temporal backbone model produces the logit vectors and then a softmax function is used to calculate the predicted probabilities as $P(x) = [p_1(x), ..., p_L(x)]$. Inspired by the findings of Hinton et al.~\cite{hinton2015distilling}, both student $P^{S}(x)$ and teacher $P^{T}(x)$ probabilities are scaled by a temperature factor for better distillation. 

Most KD literature~\cite{kim2021self, ding2022free} define the Kullback-Leibler divergence as the distillation loss.
We instead use the truncated mean squared error (MSE) $\mathcal{L}_{T-MSE}$ over the frame-wise student and teacher log probabilities as it has been shown to generate smoother predictions \cite{mstcn}. 
\begin{equation}
    \mathcal{L}_{T-MSE} = \frac{1}{LC}\sum_{l,c}\tilde{\Delta}^{2}_{l,c} \text{, where } \tilde{\Delta}_{l,c}= \begin{cases}
\Delta_{l,c} \quad: \Delta_{l,c}\le \tau\\
\tau \qquad: otherwise
\end{cases}
\end{equation}
where $\tau$ is the truncated threshold and we set it to $8.0$. In our method, we define the teacher model as the best student model from the training history, which indicates that the teacher model dynamically evolves itself as the training proceeds. Let $P^{S}_{t}(x)$ be the predictions from the student model at $t$-th epoch, the MSE loss $\Delta_{l,c}$ is defined as:
\begin{equation}
    \Delta_{l,c} = \lvert \log p^{S}_{t}(x_{l,c}) - \log p^{S}_{i<t}(x_{l-1,c}) \rvert
\end{equation}
where $i$ is the preceding epoch with the highest accuracy. It is worth noting that gradients are only calculated for student predictions $p^{S}(x_{l,c})$, while teacher predictions $p^{T}(x_{l-1,c})$ are not considered as model parameters. Combining with a cross-entropy loss $\mathcal{L}_{CE}$, decoder self-KD loss $\mathcal{L}^{D}_{self-KD}$ is defined as:
\begin{equation}
    \mathcal{L}^{D}_{self-KD} = \mathcal{L}_{CE}+ \lambda \mathcal{L}_{T-MSE}
\end{equation}
where $\lambda$ is a coefficient to determine the contribution of $\mathcal{L}_{T-MSE}$ loss. 
%......................................................%

\section{Experiments and Results}
\subsection{Experimental Settings}
\textbf{Dataset}: We evaluated our proposed method on the public Cholec80 dataset\footnote{http://camma.u-strasbg.fr/datasets}. This dataset contains 80 laparoscopic cholecystectomy procedures. Each video frame was labeled with seven phases. We used the first 40 videos as training and the rest for testing as suggested by~\cite{cholec80}. Videos were downsampled to 1 FPS. Model performance was measured using five common metrics for surgical phase recognition, namely accuracy, precision, recall, F1-score and jaccard index. Metrics were obtained from the epoch with lowest loss at decoder training. We reported the average of metrics obtained per video with corresponding standard deviation. Moreover, all our models were trained and evaluated in online mode, which means no future information was used for current prediction.

\noindent
\textbf{Encoders Training details}: We choose ResNet50~\cite{resnet} from Pytorch repo as the backbone model for baseline and self-KD experiments, with a batch size of 64 throughout 100 epochs. We followed image augmentations operations in~\cite{byol}. We used SGD to optimize the model (learning rate $1e^{-2}$ and weight decay of $1e^{-5}$). For EMA update, decay weight $\tau_0$ started at 0.9995 and was progressively updated to a value of 1, following equation $\tau_{i+1} = 1 - \frac{1}{2}(1 - \tau_0)(\cos(\frac{\pi i}{K}) + 1)$, where $i$ is current training step, and $K$ is the total number of steps. The teacher model was updated after each optimization step performed over the student model.

\noindent
\textbf{Decoders Training details}: 
For self-KD decoder training, we extracted the 2048-dim features using the aforementioned encoder and selected four popular state-of-the-art decoder backbones, including TCN, TeCNO, Trans-SVNet and GRU. Models were trained for 30 epochs using the Adam optimizer. The learning rate for TCN, TeCNO and GRU was set to 1e-3. Trans-SVNet is a three-stage model based on TeCNO. We trained it for 10 epochs with learning rate 1e-4. The self-KD loss weight for all models, $\lambda$, was set to $0.3$.\\
Our framework was implemented in PyToch 1.12.1 and all models were trained on a Tesla V100 DGXS 32GB GPU.

\subsection{Comparison with the SOTA}
\autoref{table_SOTA} presents the performance results of our framework compared with state-of-the-art. Our best model (self-KD GRU) achieved competitive results compared to MTRCNet-CL and TeCNO despite the fact that both of these models are multi-task and benefit from additional instrument signals, which are not always available. Both our best self-KD GRU model and self-KD TCN model surpassed~\cite{intuitive_paper}, despite relying on a 2D encoder instead of the more complex 3D Swin-Transformer encoder. Trans-SVNet re-implemented TeCNO to only rely on phase signals. Performance metrics reported in Trans-SVNet are higher than the metrics we achieved using our implementation based on their public code. We found out that authors of Trans-SVNet considered class imbalance during metric calculation\footnote{https://github.com/YuemingJin/TMRNet}, which is not common in the literature. We therefore assessed both Trans-SVNet and TeCNO without weighting them based on class imbalance, which are indicated by star in \autoref{table_SOTA}. Our self-KD TeCNO and self-KD Trans-SVNet exceeded their corresponding baselines (see detailed ablative experiments in~\autoref{sec:ab1}). In addition, our best model self-KD GRU model reached the new SOTA results for surgical phase recognition task. It outperformed the 3D encoder based Swin+GRU by $+2.35\%$ accuracy, $+2.39\%$ precision, and $+1.28\%$ recall.

\begin{table}[th!]
    \centering
    \caption{Performance comparison against SOTA approaches, results are shown in percentage. TeCNO* and Trans-SVNet* are two models we evaluated without taking class imbalance into account.}
    \label{table_SOTA}
    \begin{tabular}{llllll}
        \hline
         & \textbf{Acc.} & \textbf{Pre.} & \textbf{Recall} & \textbf{F1-score} & \textbf{Jaccard} \\
        \hline
        MTRCNet-CL~\cite{jin2020multi}& 89.2 & 86.9 & 88.0 & \quad -& \quad -\\
         TeCNO~\cite{tecno} & 88.56 & 81.64 &85.24& \quad- & \quad-\\
        Swin+GRU~\cite{intuitive_paper} & 90.88 & 85.07 & 85.59& \quad -& \quad -\\
        TeCNO from~\cite{transsvnet}& 88.6 & 86.5 & 87.6 & \quad-& 75.1\\
        Trans-SVNet~\cite{transsvnet} & 90.3 & 90.7 & 88.8 & \quad- & 79.3\\
        \hline
        TeCNO* & 90.20 & 84.12 & 85.34 & 82.65 & 74.07\\
        Trans-SVNet* & 89.85 & 84.96 & 84.48 & 82.25& 74.01\\
        \hline
        Self-KD TeCNO & 91.37 & 86.41 & 84.09 & 84.09  & 76.04 \\
        Self-KD Trans-SVNet & 91.68  & 86.12 & 86.06  & 84.42 & 76.56 \\
        Self-KD TCN &91.78  & 85.88  & \textbf{87.39}  &85.08  & 77.67 \\
        Self-KD GRU & \textbf{93.24}  & \textbf{87.46} & 86.87 & \textbf{85.61}  & \textbf{78.56} \\
        \hline
    \end{tabular}
\end{table}

\section{Discussion}
\subsection{Self-KD at different stages}~\label{sec:ab1}
\begin{figure}[tb!]
    \centering
    \includegraphics[trim={1cm 1cm 1cm 1cm}, width=0.9\textwidth]{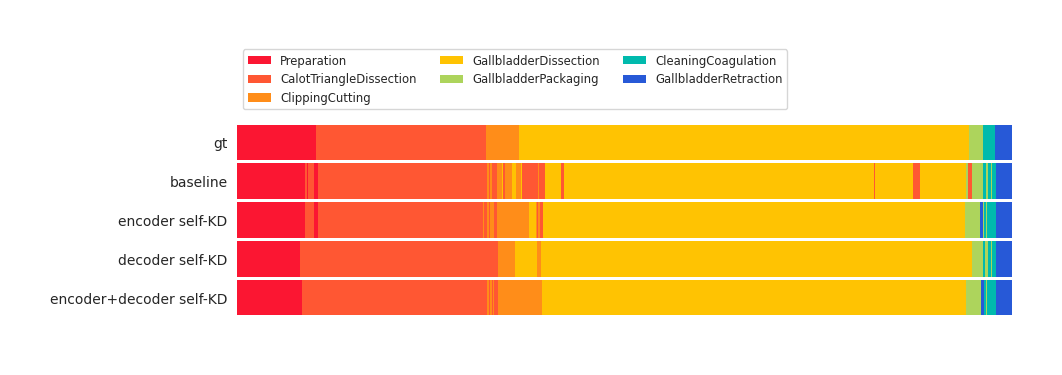}
    \caption{Qualitative results. Predictions of a ResNet50+GRU with self-knowledge distillation (self-KD) at different encoder and decoder stages are shown against the ground truth on video 61.}
    \label{fig:bar}
\end{figure}

To further investigate the efficiency of each sub-component in our framework, we conduct ablative studies with four SOTA models as follows: (1) baseline encoder + baseline decoder; (2) encoder self-KD + baseline decoder; (3) baseline encoder + decoder self-KD; (4) encoder self-KD + decoder self-KD. The results are presented in~\autoref{table_skd}, from which we observe that:\\
\textbf{(1) v.s. (2) and (1) v.s.(3):} These experiments show the impact of applying self-KD at different stages. One can notice that applying self-KD to each stage alone improved the results. In general, we notice that adding self-KD to encoder results in higher boost in overall performance. This highlights the importance of self-KD in building more robust features.\\
\textbf{(2) v.s. (4) and (3) v.s.(4):} In general, model performance benefits from two-stage self-KD (both encoder and decoder stage), especially the accuracy. For other evaluation metrics, two-stage self-KD TCN and self-KD GRU still outperform the single-stage model. We have noticed that encoder self-KD contributes more to TeCNO and Trans-SVNet in recall, F1-score, and jaccard. Because the learning ability of self-KD relies on the solution space of the model itself. TeCNO and Trans-SVNet (TeCNO based) present a small temporal convolutional based model (117K parameters) compare with reported TCN (297K parameters parameters) or GRU (481K parameters). In this regard, the temporal model has limited capacity to distillate the knowledge.\\
\textbf{(1) v.s. (4):} Self-KD can be combined with various backbone models and it has consistently improved the results over baseline models. This highlights the effectiveness of self-KD in building more robust models. Specifically our best model, self-KD GRU, not only exceeds the performance of other three models, but boosted performance over its own baseline by +5.32\% jaccard.\\
\begin{table}[hbt!]
    \centering
    \caption{Results of adding self-KD into training of architecture stages progressively, shown in percentage, for \textit{Cholec80} test set. Metrics are obtained first per video and then average and standard deviation are reported. \textit{TeCNO} and \textit{Trans-SVNet} baselines results are obtained using our own implementation for fair comparison.}
    \label{table_skd}
    \begin{adjustbox}{width=1\textwidth}
        \small
        \begin{tabular}{lccccccc}
            \hline
             & self-KD-Enc & self-KD-Dec & \textbf{Acc} & \textbf{Pre} & \textbf{Rec} & \textbf{F1-score} & \textbf{JA} \\
            \hline
            \multirow{4}{4em}{TCN} & & & 89.95±6.10 & 84.43±8.54 & 83.48±7.70 & 81.26±9.09 & 72.66±11.01\\
             & \checkmark & & 91.20±5.39 & 85.85±7.45 & 86.71±7.14 & 84.50±7.64 & 76.87±9.28\\
             & & \checkmark & 90.54±5.73 & 85.39±7.97 & 84.75±7.22 & 82.75±8.57 & 74.58±10.03\\
             & \checkmark & \checkmark & \textbf{91.78±5.20} & \textbf{85.88±7.86} &	\textbf{87.39±7.16} & \textbf{85.08±7.95} & \textbf{77.67±9.69}\\
            \hline
            \multirow{4}{4em}{TeCNO} & & & 90.20±5.76 & 84.12±8.89 & 85.34±7.18 & 82.65±8.76 & 74.07±10.85\\
             & \checkmark & & 90.78±6.32 & 84.85±8.34 & \textbf{87.27±7.11} & \textbf{84.24±8.21} & \textbf{76.53±10.28}\\
             & & \checkmark & 90.32±5.95 & 84.24±8.76 & 85.17±6.96 & 82.63±8.66 & 74.03±10.72 \\
             & \checkmark & \checkmark & \textbf{91.37±5.09} & \textbf{86.41±7.67} & 85.63±7.59 & 84.09±8.01 & 76.04±10.05 \\
            \hline
            \multirow{4}{4em}{Trans-SVNet} & & & 89.85±6.02 & 84.96±8.21 & 84.48±7.77 & 82.25±9.02 & 74.01±10.83\\
             & \checkmark & & 91.33±5.97 & 85.22±7.83 & \textbf{87.29±6.38} & \textbf{84.66±7.43} & \textbf{77.09±8.99}\\
             & & \checkmark & 90.81±5.59 & \textbf{86.93±7.59} & 85.97±6.65 & 84.34±8.06 & 76.32±10.00 \\
             & \checkmark & \checkmark & \textbf{91.68±4.79} & 86.12±7.28 & 86.06±6.94 & 84.42±7.27 &76.56 ±8.74 \\
            \hline
            \multirow{4}{4em}{GRU} & & & 89.91±7.42 & 83.10±9.45 & 84.06±8.92 & 81.66±10.02 & 73.24±12.06\\
             & \checkmark & & 92.73±5.34 & 86.04±7.47 & 86.80±7.77 & 84.75±7.75 & 77.65±9.45\\
             & & \checkmark & 91.05±5.54 & 85.28±8.88 &	82.85±7.00 & 82.13±8.05 & 73.53±10.03\\
             & \checkmark & \checkmark & \textbf{93.24±4.36}& \textbf{87.46±7.44}&	\textbf{86.87±7.46}&	\textbf{85.61±7.37}&	\textbf{78.56±9.04}  \\
            \hline
        \end{tabular}
    \end{adjustbox}
\end{table}Moreover, we show a qualitative comparison of our self-KD GRU model in \autoref{fig:bar}. It can be observed that GRU baseline model mis-classified more frames than self-KD models and suffer from severe over-segmentation issues. Adding only the encoder self-KD generates more accurate predictions but over-segments some frames around \textit{Clipping and Cutting}. Due to the limited feature representative ability of baseline ResNet, self-KD decoder model has smoother predictions but still some noisy prediction towards the end of the video. Combining the encoder and decoder self-KD, our method produces more reliable and smoother predictions. Both quantitative and qualitative results shows the superior performance of our proposed self-KD strategy and the indispensability of encoder and decoder self-KD in building robust models.

\subsection{Training data}\label{sec:ab2}
We conduct experiments to assess the effect of reducing training data sizes. We randomly removed videos from training set and compared with the baselines trained on full set. Videos are removed from the training set in two ways: excluding the video from the training altogether or excluding phase ground truth labels, in other words only self-KD loss is applied over those videos. The ResNet50+GRU model is used for the experiment reported in  \autoref{tab:low_data}. We refer to these experiments as: (1) models trained on full dataset; (2) baseline model trained on reduced sets of training videos; (3) self-KD GRU trained on reduced sets of training videos; (4) self-KD GRU trained on reduced sets of training videos in classification loss.\\ 
\noindent
\textbf{(1) v.s. (2) v.s.(3):} In general reducing the size of the training set has degraded performance for all models. We however noticed the self-KD model trained on 87.5\% of the data or even when it is trained on 75\% of the data outperformed the baseline model trained on the full dataset. This finding is important for surgical phase recognition task, as generating annotations for surgical videos can be time consuming and more expensive than in general domain computer vision dataset.\\
\textbf{(1) v.s. (4) and (3) v.s.(4):} Less reduction in performance in (4) compared with (3), highlights the ability of the proposed framework in learning from unlabeled videos. We have observed that even reducing the size of the training set by 50\%, our self-KD achieved performance on par with the same model trained on full training sets. This show the impact of using training data more effectively to generate more reliable models and hopefully paving the way towards removing the need for generating huge labeled phase datasets.  

\begin{table}[htb!]
    \centering
    \caption{Effect of reducing the number of training videos. Results are shown in percentage for Cholec80 test set with 40 training videos.}
    \label{tab:low_data}
    \begin{adjustbox}{width=0.9\textwidth}
        \small
        \begin{tabular}{ccccccc}
        \hline
            Encoder & \# videos excluded & excluded from & Accuracy & F1-score & Jaccard \\
        \hline
            ResNet50 & 0 & & 89.91±7.42 & 81.66±10.02 & 72.66±12.06  \\
            self-KD-ResNet50 & 0 & & \textbf{93.24±4.36} & 85.61±7.37 &	\textbf{78.56±9.04}  \\
        \hline
             & 5 & & 87.98±9.59 & 78.74±1.57 & 69.82±12.40 \\
            ResNet50 & 10 & training &  87.32±8.23 & 79.13±8.68 & 69.60±10.59  \\
             & 20 & & 84.31±11.35 & 74.39±9.42 & 63.86±10.79 \\
        \hline
             & 5  & & 92.38±4.32 & \textbf{85.82±8.47} & 76.33±9.80 \\
             self-KD-ResNet50 & 10 & training & 89.45±6.78 & 84.42±8.48 & 72.78±9.95\\
             & 20 & & 88.22±8.95 & 79.76±9.16 & 71.33 ± 10.63\\
        \hline
             & 5 & & 92.44±4.82  & 83.39± 8.11& 76.20±9.31\\
            self-KD-ResNet50 & 10 & classification loss & 89.99±9.10 & 82.54±8.48 & 74.26±10.67\\
             & 20 & & 89.69±7.68 & 81.99±9.07 & 74.74±10.62\\
        \hline
        \end{tabular}
    \end{adjustbox}
\end{table}

\section{Conclusion}
In this paper, we propose a self-knowledge distillation framework for surgical phase recognition. Rather than increasing model complexity, we embed the self-KD into existing models to utilize training data more effectively. Implicit knowledge is distilled in a single training process such that the encoder generates informative representations and the over-segmentation errors are reduced on the temporal decoder stage. We assessed our framework by applying it to four popular surgical phase recognition models. Our experiments show that our framework outperforms baselines over all evaluation metrics and our best model self-KD GRU achieves new SOTA performance We believe the results can answer our proposed question: rather than always going to a more complex solution, training data can be utilized efficiently by our self-knowledge distillation framework in building more robust models.Due to the limitation of small dataset size, our future work will be extended to larger datasets to better study the impact of our framework.
\backmatter

%\bmhead{Supplementary information}

%\bmhead{Acknowledgements}
\section*{Declarations}

%%Some journals require declarations to be submitted in a standardised format. Please check the Instructions for Authors of the journal to which you are submitting to see if you need to complete this section. If yes, your manuscript must contain the following sections under the heading `Declarations':

\begin{itemize}
%%\item Funding
\item Funding: This work was funded by Medtronic plc. 
\item Competing interests: Mr. Barbarisi, Drs. Zhang, Kadkhodamohammadi and Luengo; and Prof. Stoyanov are employees of Medtronic plc. Prof. Stoyanov is also a co-founder and shareholder in Odin Vision, Ltd.
\item Ethics and consent: This article does not contain any studies with human participants or animals performed by any of the authors.
\item Data, code and/or material availability: Privately held

%%\item Consent to participate
%%\item Consent for publication
%%\item Availability of data and materials
%%\item Code availability 
%%\item Authors' contributions
\end{itemize}

%%\noindent
%%If any of the sections are not relevant to your manuscript, please include the heading and write `Not applicable' for that section. 

%%\begin{appendices}

%%\section{Appendix}\label{secA1}



%% \end{appendices}

\bibliography{sn-bibliography}% common bib file


\end{document}